\documentstyle[twoside]{article}

\catcode`\@=11
\long\def\@makefntext#1{
\protect\noindent \hbox to 3.2pt {\hskip-.9pt  
$^{{\eightrm\@thefnmark}}$\hfil}#1\hfill}		

\def\@makefnmark{\hbox to 0pt{$^{\@thefnmark}$\hss}}	
	
\def\ps@myheadings{\let\@mkboth\@gobbletwo
\def\@oddhead{\hbox{}
\rightmark\hfil\eightrm\thepage}   
\def\@oddfoot{}\def\@evenhead{\eightrm\thepage\hfil
\leftmark\hbox{}}\def\@evenfoot{}
\def\sectionmark##1{}\def\subsectionmark##1{}}

\oddsidemargin=\evensidemargin
\newcounter{sectionc}\newcounter{subsectionc}\newcounter{subsubsectionc}
\renewcommand{\section}[1] {\vspace{12pt}\addtocounter{sectionc}{1} 
\setcounter{subsectionc}{0}\setcounter{subsubsectionc}{0}\noindent 
	{\tenbf\thesectionc. #1}\par\vspace{5pt}}
\renewcommand{\subsection}[1] {\vspace{12pt}\addtocounter{subsectionc}{1} 
	\setcounter{subsubsectionc}{0}\noindent 
	{\bf\thesectionc.\thesubsectionc. {\kern1pt \bfit #1}}\par\vspace{5pt}}
\renewcommand{\subsubsection}[1] {\vspace{12pt}\addtocounter{subsubsectionc}{1}
	\noindent{\tenrm\thesectionc.\thesubsectionc.\thesubsubsectionc.
	{\kern1pt \tenit #1}}\par\vspace{5pt}}
\newcommand{\nonumsection}[1] {\vspace{12pt}\noindent{\tenbf #1}
	\par\vspace{5pt}}

\newcounter{appendixc}
\newcounter{subappendixc}[appendixc]
\newcounter{subsubappendixc}[subappendixc]
\renewcommand{\thesubappendixc}{\Alph{appendixc}.\arabic{subappendixc}}
\renewcommand{\thesubsubappendixc}
	{\Alph{appendixc}.\arabic{subappendixc}.\arabic{subsubappendixc}}

\renewcommand{\appendix}[1] {\vspace{12pt}
        \refstepcounter{appendixc}
        \setcounter{figure}{0}
        \setcounter{table}{0}
        \setcounter{lemma}{0}
        \setcounter{theorem}{0}
        \setcounter{corollary}{0}
        \setcounter{definition}{0}
        \setcounter{equation}{0}
        \renewcommand{\thefigure}{\Alph{appendixc}.\arabic{figure}}
        \renewcommand{\thetable}{\Alph{appendixc}.\arabic{table}}
        \renewcommand{\theappendixc}{\Alph{appendixc}}
        \renewcommand{\thelemma}{\Alph{appendixc}.\arabic{lemma}}
        \renewcommand{\thetheorem}{\Alph{appendixc}.\arabic{theorem}}
        \renewcommand{\thedefinition}{\Alph{appendixc}.\arabic{definition}}
        \renewcommand{\thecorollary}{\Alph{appendixc}.\arabic{corollary}}
        \renewcommand{\theequation}{\Alph{appendixc}.\arabic{equation}}
        \noindent{\tenbf Appendix \theappendixc #1}\par\vspace{5pt}}
\newcommand{\subappendix}[1] {\vspace{12pt}
        \refstepcounter{subappendixc}
        \noindent{\bf Appendix \thesubappendixc. {\kern1pt \bfit #1}}
	\par\vspace{5pt}}
\newcommand{\subsubappendix}[1] {\vspace{12pt}
        \refstepcounter{subsubappendixc}
        \noindent{\rm Appendix \thesubsubappendixc. {\kern1pt \tenit #1}}
	\par\vspace{5pt}}

\newcommand{\textlineskip}{\baselineskip=13pt}
\newcommand{\smalllineskip}{\baselineskip=10pt}

\def\eightcirc{
\begin{picture}(0,0)
\put(4.4,1.8){\circle{6.5}}
\end{picture}}
\def\eightcopyright{\eightcirc\kern2.7pt\hbox{\eightrm c}} 

\newcommand{\copyrightheading}[1]
	{\vspace*{-2.5cm}\smalllineskip{\flushleft
	{\footnotesize International Journal of Modern Physics A, #1}\\
	{\footnotesize $\eightcopyright$\, World Scientific Publishing
	 Company}\\
	 }}

\def\abstracts#1#2#3{{
	\centering{\begin{minipage}{4.5in}\baselineskip=10pt\footnotesize
	\parindent=0pt #1\par 
	\parindent=15pt #2\par
	\parindent=15pt #3
	\end{minipage}}\par}} 

\renewenvironment{thebibliography}[1]
	{\frenchspacing
	 \ninerm\baselineskip=11pt
	 \begin{list}{\arabic{enumi}.}
	{\usecounter{enumi}\setlength{\parsep}{0pt}
	 \setlength{\leftmargin 12.7pt}{\rightmargin 0pt} 
	 \setlength{\itemsep}{0pt} \settowidth
	{\labelwidth}{#1.}\sloppy}}{\end{list}}

\newcounter{itemlistc}
\newcounter{romanlistc}
\newcounter{alphlistc}
\newcounter{arabiclistc}

\newcommand{\fcaption}[1]{
        \refstepcounter{figure}
        \setbox\@tempboxa = \hbox{\footnotesize Fig.~\thefigure. #1}
        \ifdim \wd\@tempboxa > 5in
           {\begin{center}
        \parbox{5in}{\footnotesize\smalllineskip Fig.~\thefigure. #1}
            \end{center}}
        \else
             {\begin{center}
             {\footnotesize Fig.~\thefigure. #1}
              \end{center}}
        \fi}

\newcommand{\tcaption}[1]{
        \refstepcounter{table}
        \setbox\@tempboxa = \hbox{\footnotesize Table~\thetable. #1}
        \ifdim \wd\@tempboxa > 5in
           {\begin{center}
        \parbox{5in}{\footnotesize\smalllineskip Table~\thetable. #1}
            \end{center}}
        \else
             {\begin{center}
             {\footnotesize Table~\thetable. #1}
              \end{center}}
        \fi}

\def\@citex[#1]#2{\if@filesw\immediate\write\@auxout
	{\string\citation{#2}}\fi
\def\@citea{}\@cite{\@for\@citeb:=#2\do
	{\@citea\def\@citea{,}\@ifundefined
	{b@\@citeb}{{\bf ?}\@warning
	{Citation `\@citeb' on page \thepage \space undefined}}
	{\csname b@\@citeb\endcsname}}}{#1}}

\newif\if@cghi
\def\cite{\@cghitrue\@ifnextchar [{\@tempswatrue
	\@citex}{\@tempswafalse\@citex[]}}
\def\citelow{\@cghifalse\@ifnextchar [{\@tempswatrue
	\@citex}{\@tempswafalse\@citex[]}}
\def\@cite#1#2{{$\null^{#1}$\if@tempswa\typeout
	{IJCGA warning: optional citation argument 
	ignored: `#2'} \fi}}

\def\pmb#1{\setbox0=\hbox{#1}
	\kern-.025em\copy0\kern-\wd0
	\kern.05em\copy0\kern-\wd0
	\kern-.025em\raise.0433em\box0}

\def\fnt#1#2{\footnotetext{\kern-.3em
	{$^{\mbox{\scriptsize #1}}$}{#2}}}

\def\fpage#1{\begingroup
\voffset=.3in
\thispagestyle{empty}\begin{table}[b]\centerline{\footnotesize #1}
	\end{table}\endgroup}

\font\tenrm=cmr10
\font\tenit=cmti10 
\font\tenbf=cmbx10
\font\bfit=cmbxti10 at 10pt
\font\ninerm=cmr9

\font\eightrm=cmr8

\def\qed{\hbox{${\vcenter{\vbox{			
   \hrule height 0.4pt\hbox{\vrule width 0.4pt height 6pt
   \kern5pt\vrule width 0.4pt}\hrule height 0.4pt}}}$}}

\begin{document}


\normalsize\textlineskip
\thispagestyle{empty}
\setcounter{page}{1}

\copyrightheading{}			

\vspace*{0.88truein}

\fpage{1}
\centerline{\bf BOSONIC TOPCOLOR
\footnote{Talk given by A. Aranda at DPF 2000, Ohio State, August 2000. 
William and Mary preprint WM-00-112.}}
\vspace*{0.035truein}
\vspace*{0.37truein}
\centerline{Alfredo Aranda and Christopher D. Carone}
\vspace*{0.015truein}
\centerline{\footnotesize\it Nuclear and Particle Theory Group}
\baselineskip=10pt
\centerline{\footnotesize\it Physics Department, College of William 
and Mary}
\baselineskip=10pt
\centerline{\footnotesize\it Williamsburg, VA 23187-8795, USA}
\vspace*{0.225truein}

\vspace*{0.21truein}
\abstracts{A topcolor model is presented that contains both
composite and fundamental scalar fields.  Strong dynamics accounts
for most of the top quark mass and part of the electroweak symmetry
breaking scale. The fundamental scalar is weakly coupled and
transmits its share of electroweak symmetry breaking to the 
light fermions. The model is allowed by the current experimental bounds, 
and can give a potentially large contribution to $D^0-\bar{D^0}$ mixing.}{}{}
\textlineskip			
\vspace*{12pt}			

\vspace*{-0.5pt}
\noindent
In this talk we present a topcolor model that contains two scalar 
fields: a composite scalar generated by strong dynamics at a 
scale $\Lambda$, and a fundamental, weakly-coupled scalar.\cite{aracar} 
For simplicity, we assume that the strong dynamics affect only the fields 
$t_R$ and $\Psi_L = (t_L\,,\,b_L)$. The role of the fundamental scalar 
is to transmit electroweak symmetry breaking (EWSB) to the light 
fermions.

In most conventional models of dynamical electroweak symmetry
breaking, light fermion masses are achieved by introducing
higher-dimension operators, suppressed by the cut off of the theory.
A complete set of such operators generically contains some that
contribute to flavor changing neutral current (FCNC) processes at  
unacceptable levels.  One may hope to make progress by exploring
models in which these operators have a renormalizable origin.
However, as extended technicolor models demonstrate, the problem
often persists.  FCNC operators can be  sufficiently 
suppressed if they are proportional to the product of small,
off-diagonal Yukawa couplings, as in models where such operators
are generated by the exchange of a fundamental scalar 
field with the quantum numbers of the standard model Higgs 
boson.\cite{carsim}.  If this scalar is integrated out of the theory, 
one obtains a conventional topcolor model\cite{hill}, augmented by 
an acceptable set of higher-dimension operators, including some which 
generate light fermion masses.  If, on the other hand, the fundamental 
scalar field remains in the low-energy theory, one obtains the model we 
call bosonic topcolor above.\cite{aracar,also}   An advantage of bosonic 
topcolor is that the fundamental scalar may share significantly in the 
breaking of electroweak symmetry, so that the usual problematic 
relation\cite{bhlym} between the dynamical top mass and the electroweak 
scale can be successfully avoided.

We note that the scenario described above may not be unnatural in the 
context of large extra dimensions and low-scale quantum 
gravity.\cite{extra}  We see three reasons why this may be the 
case: (i) Fundamental scalars arise in string theory.  (ii)  The 
Planck scale can be low, so that the fine-tuning required to keep the 
fundamental scalar light is no greater than that associated with the 
composite scalar.  (iii)  Exchange of a nonperturbatively large number of 
gluon Kaluza-Klein (KK) excitations may provide the source of strong 
dynamics~\cite{dobrescu}, while KK excitations of the fundamental scalar 
(if it is a bulk field) couple non-universally and might provide a means 
for achieving the desired vacuum tilting.  Below we study the simplest 
(four-dimensional) effective theory, which assumes only (i) and (ii) above.  
We expect the phenomenology of this model to be similar to more 
realistic examples yet to be explored.

Our high-energy theory is defined by
\begin{eqnarray}
{\cal L}_{\rm H} & = & D_\mu H^\dagger D^\mu H - m^2_H H^\dagger H
- \lambda (H^\dagger H)^2  \nonumber \\ 
                 &   & - h_t (\overline{\Psi}_L H t_R + h.c.)
+ \frac{\kappa}{\Lambda^2} \overline{\Psi}_L t_R \overline{t}_R \Psi_L \,\,,
\label{eq:L}
\end{eqnarray}
where $H$ is the fundamental scalar, and the last interaction term is 
of NJL type.  As mentioned before, we assume $H$ couples weakly to all
fermions, so that, in particular, $h_t$ is small. 

Just below $\Lambda$, we rewrite Eq.~(\ref{eq:L}) 
\begin{eqnarray} \label{lagrangian}
\nonumber
{\cal{L}} & = & D_{\mu} H^{\dagger} D^{\mu} H 
- m_H^{2} H^{\dagger} H  - \lambda \left( 
H^{\dagger} H \right)^{2} - c \Lambda^{2} \Sigma^{\dagger} \Sigma
 \\
          &   & - h_{t} ( \bar{\Psi}_{L} t_{R} H + h.c.) -
g_t (\bar{\Psi}_{L} t_{R} \Sigma + h.c.)\,\, ,
\end{eqnarray}
where $\Sigma =-g_t(\bar{t}_R\psi_L)/(c\Lambda^2)$ is a non-propagating
auxiliary field. At scales $\mu < \Lambda$, quantum corrections induce 
a kinetic term for $\Sigma$ so that it becomes dynamical.  The 
low-energy theory is therefore a very nongeneric two-Higgs-doublet model 
of type III. Study of the scalar potential at the scale $\mu$ reveals that 
both scalar fields can acquire vacuum expectation values (vevs) and 
participate in EWSB, even when $m^2_H >0$.  In this case EWSB is clearly 
triggered by the strong dynamics -- no negative squared mass 
term is introduced by hand.

For a fixed choice of the scales $\mu$, and $\Lambda$, the phenomenology of 
this model may be conveniently parameterized in terms of the couplings $h_t$ 
and $\lambda$, as well as the ratio of Higgs vevs, $\tan\beta$.  Setting
$\Lambda \mbox{\raisebox{-1.0ex} 
{$\stackrel{\textstyle ~<~}{\textstyle \sim}$}} 100$~TeV, and identifying 
$\mu$ with the electroweak scale, allowed regions of the $h_t$-$\lambda$ 
plane are presented in Ref.~\cite{aracar}, for $\tan\beta \approx 1$.  This
analysis takes into account the bounds from Higgs boson searches, precision 
electroweak parameters and FCNC processes.  As an example, setting 
$\Lambda = 10$~TeV, we find an allowed region in the 
range $\,\,$ $0.1 < \lambda < 5.0$ $\,\,$ for $\,\,$ 
$3.2 \times 10^{-3} < h_t < 1.6 \times 10^{-2}$.   The range in $h_t$ is 
delimited from above by the bound from the $T$ parameter, and below from 
the $b \rightarrow s \gamma$ branching fraction.  For $\lambda > 5.0$, the 
lower bound on $h_t$ comes from the $S$ parameter, and reduces the allowed 
region almost linearly from the range quoted above to $ 1.3 \times 10^{-2}
< h_t < 1.9 \times 10^{-2}$ at $\lambda = 10.0$.  Viable parameter
ranges can be obtained for other choices of $\Lambda$ as well.

Since the composite field $\Sigma$ couples only to third generation, there 
is a potentially distinctive source of flavor violation in 
the model. Consider the mass matrix for the up-type quarks,
\begin{eqnarray}\label{mu}
M^U = Y^U_\Sigma \frac{v_1}{\sqrt{2}} +  Y^U_H \frac{v_2}{\sqrt{2}} \,\,\, ,
\end{eqnarray}
where the matrices $Y$ are Yukawa couplings, and the vevs of $\Sigma$ and $H$
are given by  $v_1/\sqrt{2}$ and $v_2/\sqrt{2}$, respectively.  Viable 
forms for the Yukawa textures are given by
\begin{equation}
M^U = \left(\begin{array}{ccc}0 &0&0\\ 0 & 0 & 0 \\ 0 & 0 & r g_t \end{array}
\right)\frac{v_1}{\sqrt{2}} + \left(\begin{array}{ccc} \lambda^8 & \lambda^5 &
\lambda^3 \\ \lambda^5 & \lambda^4 & \lambda^2 \\ \lambda^3 & \lambda^2 & h_t
\end{array}\right)\frac{v_2}{\sqrt{2}} \,\,\, ,
\end{equation}
where $\lambda=0.22$ is the Cabibbo angle, and $r$ is a rescaling that
results from placing the $\Sigma$ kinetic terms in canonical form.  Notice 
that these matrices approximately reproduce the correct up quark masses and 
CKM angles.  Performing CKM-like rotations to diagonalize $M^U$  yields 
$1$-$2$ neutral scalar couplings of order $\lambda^5$.  Thus, there is a 
contribution to $D^0$-$\overline{D^0}$ mixing,
\begin{equation}
|\frac{\Delta m_D}{m_D}|_{{\rm new}} 
\approx \lambda^{10} \frac{f_D^2}{12 M_\phi^2}\left[
-1+11\frac{m_D^2}{(m_c+m_u)^2}\right]  \,\,\, .
\label{eq:ddbar}
\end{equation}
For $f_D\approx 200$~MeV, this contribution saturates the current experimental
bound, $\Delta m_D < 1.58\times 10^{-10}$~MeV, for 
$M_\phi \mbox{\raisebox{-1.0ex} {$\stackrel{\textstyle ~<~}{\textstyle \sim}$}}
495$~GeV.  Of course, it is possible to construct textures such that this
effect vanishes (for example, if the CKM angles originate solely from the 
diagonalization of $M^D$).  Generically, however, we expect the new 
contribution to $D^0$-$\overline{D^0}$ mixing in this model to be large.

In this talk we have presented a model in which electroweak symmetry breaking 
is triggered by third generation strong dynamics and transmitted
to lighter fermions by a fundamental scalar. The model is phenomenologically 
viable, and can give a contribution to $D^0-\bar{D^0}$ mixing as large as the 
current experimental bound.

\nonumsection{Acknowledgements}
\noindent
We thank the National Science Foundation for support under Grant No.
PHY-9900657, and the Jeffress Memorial Trust for support under Grant No. J-532.


\begin{thebibliography}{000}
\bibitem{aracar}
A. Aranda and C.D. Carone, Phys. Lett. B488 (2000), 351-358.

\bibitem{carsim}
E.H. Simmons, Nucl.\ Phys.\ B312 (1989) 253; C.D. Carone and
E.H. Simmons, Nucl.\ Phys.\ B397 (1993) 591; C.D. Carone and
E.H. Simmons, Phys.\ Lett.\ B344 (1995) 287.
C.D. Carone and H. Georgi, Phys.\ Rev.\ D49 (1994) 1427.

\bibitem{hill}
C.T. Hill,  Phys.\ Lett.\ B266 (1991) 419.

\bibitem{also}
See also J. D. Wells, Phys.\ Rev.\ D56 (1997) 1504. 

\bibitem{bhlym}
W.A. Bardeen, C.T. Hill and M. Lindner, Phys.\ Rev.\ D41 (1990) 1647;
V.A. Miransky, M. Tanabashi and K. Yamawaki, Phys.\ Lett.\ B221 (1989) 177; 
Mod.\ Phys.\ Lett.\ A4 (1989) 1043.

\bibitem{extra}
N. Arkani-Hamed, S. Dimopoulos, and G. Dvali,
Phys.\ Lett.\ B429 (1998) 263; K. R. Dienes,
E. Dudas and T. Gherghetta, Phys.\ Lett.\ B436 (1998) 55.

\bibitem{dobrescu}
B. A. Dobrescu, hep-ph/9903407; H-C Cheng, B. A. Dobrescu, C. T. Hill,
hep-ph/9912343.

\end{thebibliography}
\end{document}